\documentclass[preprint, amssymb,  aps, prl]{revtex4-1}
\usepackage{graphicx, subfigure}
\usepackage{bm}
\usepackage{times}
\usepackage{amsfonts}
\usepackage{amsmath}
\usepackage{bm}

\newcommand{\phik}{\phi_k}
\newcommand{\thetak}{\theta_k}

\begin{document}

\title{Excitation of single multipolar modes with engineered cylindrically symmetric fields}

\author{Xavier Zambrana-Puyalto$^{1,2}$, Xavier Vidal$^{1}$ and Gabriel Molina-Terriza$^{1,2}$}

\affiliation{$^{1}$QsciTech and Department of Physics and Astronomy, Macquarie University, 2109 NSW, Australia}
\affiliation{$^{2}$ARC Center of Excellence on Engineered Quantum Systems (EQuS)}

\email{xavier.zambranapuyalto@mq.edu.au} 

\begin{abstract}
We present a new method to address multipolar resonances and to control the scattered field of a spherical scatterer. This method is based on the engineering of the multipolar content of the incident beam. We propose experimentally feasible techniques to generate light beams which contain only a few multipolar modes. The technique uses incident beams with a well defined component of the angular momentum and appropriate focusing with aplanatic lenses. The control of the multipolar content of light beams allow for the excitation of single Mie resonances and unprecedented control of the scattered field from spherical particles.  
\end{abstract}

\maketitle


\section{Introduction}
Resonances are ubiquitous in the Physical Sciences and Engineering. In a general sense, a resonance expresses that a system is efficiently coupled to an external excitation within the interval of a certain parameter. When properly controlled, this efficient coupling can be used to measure different phenomena with high sensitivity. For instance, the control of nuclear magnetic resonances in nuclei has enabled a great number of applications in biology and medicine \cite{Gadian1982}. Unfortunately, the lack of control of these resonant phenomena can derive in catastrophic results, one of the most famous being the collapse of the Takoma bridge in the 1940's \cite{Billah1991}. Recently, resonances in plasmonic systems have been gaining a lot of interest \cite{Garcia-Vidal2010} and some techniques to control them have been proposed \cite{Lopez2002, Mojarad2008, preKim2012}. Most resonances can be approximately modelled with the Lorentzian distribution. However, in 1961 a new fundamental kind of resonance was discovered by Fano \cite{Fano1961}. Although it was firstly regarded as a quantum mechanical effect, the Fano resonance is very present in optics and plasmonics. In particular, one of the most important analytical solutions of Maxwell equations in inhomogeneous media, \textit{i.e.} Mie Theory, gives rise to Fano resonances, too \cite{Fano2010}. The Mie Theory \cite{Hergert2012} or the more modern Generalized Lorenz-Mie Theory (GLMT) \cite{GLMT_book} describes the interaction between an incident electromagnetic field propagating in a lossless, homogeneous, isotropic medium and a homogeneous isotropic sphere. The number of applications of this theory is immense. The reader is referred to Gouesbet \textit{et al.} for an extensive review \cite{Gouesbet2011}. 

The most appropriate electromagnetic modes for describing the physical phenomena associated with the GLMT are the multipolar modes: $\lbrace \mathbf{A}_{jm_z}^{(m)},  \mathbf{A}_{jm_z}^{(e)} \rbrace$, where $(m)$ and $(e)$ stand for magnetic and electric multipoles. These modes are eigenvectors of the total angular momentum (AM) operator $J^2$ and one of its projections such as $J_z$ with respective values $j$ and $m_z$ \cite{Rose1955, Nora2012}. This set of modes exploit the spherical symmetries of the kind of problems studied by the GLMT. As a consequence, the electromagnetic eigenmodes of the scattered field from a spherical object are precisely the multipolar modes. Experimentally, it is very challenging to prepare and detect a single multipole with an arbitrary value of both $j$ and $m_z$. This is due to the spherical and non-paraxial character of these modes. In this paper we propose a technique to control multipolar resonances of dielectric spheres by engineering the multipolar content of experimentally realizable light beams. In our approach, we use focused cylindrically symmetric beams. These light beams possess a well defined value of $J_z$. In 1992 Allen and co-workers showed that in the paraxial approximation one could find a set of modes which had a well defined value of the AM \cite{Allen1992}. A particular set of paraxial modes with this property are the Laguerre-Gaussian modes (LG$_{q,l}$, with $q$ the radial index and $l$ the azimuthal one). This important finding opened up a whole new field, where the ability to independently control the AM and the polarization components of paraxial beams has allowed innumerable applications in quantum optics, microscopy, biology, optical trapping and astrophysics, just to mention a few \cite{Sonja2008}. The total AM of a vectorial field is composed of an orbital part ($\mathbf{L}$) and a spin part ($\mathbf{S}$), \textit{i.e.} $\mathbf{J}=\mathbf{L}+\mathbf{S}$. Nevertheless, it is important to note that these two components cannot ordinarily be separated. When both of these operators are applied to a Maxwell field, the result from that operation is not a Maxwell field \cite{Lifshitz1982,preIvan2012PRA}.

This paper presents two main findings. First, we show how to experimentally control the multipolar content of cylindrically symmetric modes. We advance that by controlling a component of the AM and the focusing of the incoming beam we can engineer fields which contain just few predetermined multipolar modes. Secondly, we demonstrate that by properly engineering these modes, we can also control the scattered field as well as address single resonances of an arbitrary, homogeneous and isotropic sphere in a lossless, homogeneous and isotropic medium. The second idea falls under the GLMT and is highly relevant to applications where control of the scattered field from a spherical object is desired. This scattering problem is solved by decomposing all the fields involved in the problem (incident $\mathbf{E}^\mathrm{i}$, scattered $\mathbf{E}^\mathrm{sca}$, and interior $\mathbf{E}^\mathrm{l}$) into multipoles and then use the boundary conditions. The three fields of the problem have the same formal expression, with the only difference that the multipolar modes of the scattered field have Hankel functions for the radial amplitudes, while both interior and incident fields have Bessel functions. However, the multipolar decomposition of the fields $\mathbf{E}^\mathrm{sca}$ and $\mathbf{E}^\mathrm{l}$ have all their complex amplitudes modulated by a Mie coefficient. These Mie coefficients are obtained by applying the Maxwell boundary conditions and depend on the AM as well as the parity. Thus, the only technicality of the problem is the projection of the incident field into the basis of multipoles. 



The cylindrical symmetry of the incident beam simplifies the problem since the other two fields are required to possess a well defined value of $J_z$, too. In this way, we can solve the problem using a mixture of analytical and numerical techniques. More importantly, by exploiting the symmetries of the system we can get a further insight on the problem only using considerations of conservation of AM. Previous works considered similar problems where spherical metallic particles were excited with Laguerre-Gaussian beams \cite{vandeNes07}. There, the use of purely numerical techniques allowed the authors to distinguish certain properties of the scattered field. Our approach is different and more general. We extend the previous results to excitation with any cylindrically symmetric field and show a clear way of controlling the scattering from spherical particles. The results presented in this paper allow for a better understanding of recent experiments of scattering of silica spheres with Laguerre-Gaussian beams \cite{Gabi2012OL}, as our formulation also includes the paraxial case. In fact, they could also help to expand the computations of the optical forces when the incident beam is not a plane wave \cite{Stilgoe2008} and when the particle is not much smaller than the wavelength \cite{Nieto-Vesperinas2010, Novitsky2011}. Last but not least, our technique to control the multipolar content of cylindrically symmetric beams along with its immediate application to excite single multipolar modes could open up new possibilities in the forefront research of magnetic resonances \cite{Garcia-Etxarri2011, Kuznetsov2012, Miroshnichenko2012, Filonov2012}.

\section{Control of the multipolar content of a focused beam}

The experimental set-up which we propose is as follows. First, we prepare a monochromatic paraxial beam with a well defined $J_z$ controlling its polarization and spatial properties separately. This can be typically achieved with either spatial light modulators or holograms and a set of waveplates and polarisers. After this step, we will focus the resulting beam with a lens which we model as aplanatic \cite{Novotny2006}. The resulting focused beam interacts with the sphere and we analyse the scattering field. From now on, we will express our results in terms of the vector potential in the Coulomb gauge. All electric fields can be recovered using the typical formulas, \textit{i.e.} $\mathbf{E}=ik\mathbf{A}$ \cite{Rose1955}. Hence, the expression of the paraxial beam previously mentioned is $\mathbf{A}^{\mathrm{i}}(\rho,\phi,z=z_1)=F_l(\rho)\exp(il\phi)\exp(ikz_1)\mathbf{e}_p$, where $\{\rho,\phi,z\}$ are the spatial cylindrical coordinates, $z_1$ is the transverse plane right before the lens, $k$ is the modulus of the wavevector, $F_l$ is an arbitrary function representing the complex radial amplitude and $\mathbf{e}_p=(\hat{x}+i\,p\hat{y})/\sqrt{2}$ is a unit polarization vector representing right ($p=+1$) or left ($p=-1$) circular polarization. Here, we are disregarding terms in the amplitude of the order $1/(kd)^2$, where $d$ is the typical size of our light beam, which in our case is approximately the diameter of the focusing lens input aperture. Any paraxial beam can be decomposed as a superposition of this set of modes \cite{Gabi2001}. Note that a paraxial beam with circular polarization has, with a very good approximation, a well defined helicity $\Lambda = \mathbf{J} \cdot \mathbf{P} / \vert \mathbf{P} \vert $, where $\mathbf{P}$ and $\mathbf{J}$ are the linear and angular momentum operators respectively \cite{preIvan2012}. As mentioned before, then the paraxial beam impinges on the back of the microscope objective and it is focused. Microscope objective manufacturers make the objectives so that they behave approximately as aplanatic lenses. In addition to this, they coat the lenses in a way such that the transmission coefficients for the $\hat{\mathbf{s}}$ and $\hat{\mathbf{p}}$ polarized waves \cite{Morse1953} are equal, \textit{i.e.} $t^s(\theta)=t^p(\theta)$. Therefore, specifying the aplanatic model described in \cite{Novotny2006} for this case, and taking into account that this model preserves the helicity and the AM of the incident beam \cite{preIvan2012PRA}, we can write the field at the focus of the lens as 
\begin{eqnarray}
\label{Afoc}
& &\mathbf{A}^{\mathrm{foc}}(\mathbf{r}) =  C
\int_{0}^{2\pi} d\phi_k \int_0^{\theta_M} \sin{\thetak} d\thetak \sqrt{n \cos \thetak} \times \nonumber\\
 & & F_l(f \sin\thetak ) \exp(i (l+p) \phi_k) \mathbf{M}(\thetak,\phik)\mathbf{e}_p \exp(i \mathbf{k} \mathbf{r}) 
\end{eqnarray}
where $C=f \exp(-ikf)/2\pi$; $f$ is the focal of the lens; $\theta_M=\arcsin(\mathrm{NA})$, with NA the numerical aperture of the lens; $\mathbf{k}=k( \sin\thetak \cos\phi_k, \sin\thetak \sin\phi_k, \cos\thetak)$; $n$ is the index of refraction of the lens; $\mathbf{r}=(\rho \cos(\phi),\rho \sin(\phi),z)$ is the position vector; and $\mathbf{M}(\thetak,\phi_k)$ is the rotation matrix applied to the polarization vector so that the transversality of the field is fulfilled. Note, that field in Eq. (\ref{Afoc}) fulfils the exact Maxwell equations.

We will now describe how these beams will allow us to control the scattered field off a sphere. As previously mentioned, this problem is solved once the incident field ($\mathbf{A}^{\mathrm{foc}}$ in our case) is decomposed into multipolar modes. This decomposition can be made using the method put forward in \cite{Gabi2008}:

\small{
\renewcommand{\arraystretch}{2.5}
\begin{equation}
\begin{array}{ccl}
&&\mathbf{A}^{\mathrm{foc}}= \displaystyle\sum_{j=| l+p |}^{\infty} i^j (2j+1)^{1/2} C_{jlp}\left[ \mathbf{A}_{j(l+p)}^{(m)}+ip\mathbf{A}_{j(l+p)}^{(e)} \right]\\
&&C_{jlp}= C \displaystyle\int_0^{\theta_M}\sin\thetak d\thetak d_{(l+p)p}^j(\thetak)F_l(f \sin\thetak) \sqrt{n \cos \thetak}
\end{array}
\label{decomp}
\end{equation}
\renewcommand{\arraystretch}{1}
\normalsize
As previously stated, the lens does not break the cylindrical symmetry of the beam, meaning that after passing through the lens, the beam still carries the same AM component $J_z=m_z$. Thus, one can see in the decomposition of Eq. (\ref{decomp}) that $m_z$ is the same for all multipoles and corresponds to $m_z=l+p$, which bounds the value of $j\ge|l+p|$. The coefficients of the decomposition $C_{jlp}$ only depend on a one dimensional integral which averages the radial complex amplitude of the input beam ($F_l$) over the reduced rotation matrix $d_{mp}^j(\theta)$, which can be found in \cite{Rose1957}. Finally, note that the beam still has a well defined helicity with value $p$. Indeed, if we apply the helicity operator, which can be written as $\Lambda = (\nabla \times)/k$ in differential form \cite{Tung1985}, to $\mathbf{A}^{\mathrm{foc}}$ in Eq. (\ref{decomp}) we obtain $\Lambda \mathbf{A}^\mathrm{foc} = p \mathbf{A}^\mathrm{foc}$. This is a consequence of the following two relations \cite{Rose1955}:
\begin{equation}
\Lambda \mathbf{A}_{jm_z}^{(m)}= i \mathbf{A}_{jm_z}^{(e)}, \quad \Lambda \mathbf{A}_{jm_z}^{(e)}= -i \mathbf{A}_{jm_z}^{(m)}
\label{TM-TE}
\end{equation}

\begin{figure}[htbp]
\centering\includegraphics[width=12cm]{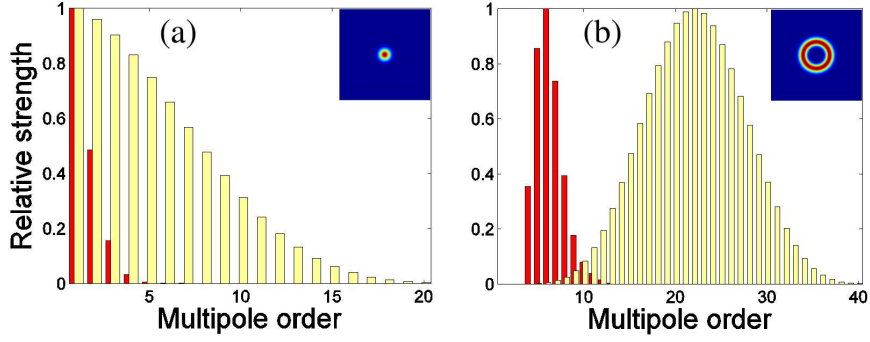}
\caption{Multipolar decomposition ($\vert C_{jlp} \vert^2$) for different cases. The insets represent the intensity plots of the modes used for each simulation. The red coloured bars indicate NA=0.25, and the blue ones NA=0.9. The multipolar decomposition of (a) LG$_{0,0}$ and (b) LG$_{0,3}$ is presented. Note that $C_{jlp} $ can be described with very few multipoles when we use a high NA microscope objective to focus the beam.}
\end{figure}
Figure 1 exemplifies one of the main results of this paper. It shows the multipolar decomposition of different focused beams. Two different input beams were used: a Gaussian beam and a Laguerre-Gaussian with optical charge $l=3$, both right circularly polarized. The width of both beams was chosen so that the entrance pupil was filled. We consider two aplanatic lenses with NA's equal to 0.25 and 0.9. Our results show that the higher the NA, the narrower the distribution of multipolar modes is. The other parameter to control the distribution of multipoles is the $J_z$ of the incident beam. As mentioned earlier, the distribution of multipolar amplitudes is zero for values of $j<|l+p|$. This allows us to control the multipolar content of an optical field.

\section{Excitation of single multipolar modes}

The decomposition in Eq. (\ref{decomp}) and the use of the results of Mie scattering are the needed elements to calculate the scattered and interior fields,
\small{
\begin{equation}
\begin{array}{l}
\mathbf{A}^\mathrm{sca}=-\displaystyle\sum_{j=\vert l+p \vert}^{\infty} i^j (2j+1)^{1/2}  C_{jlp} \left[ b_j\mathbf{A}_{j(l+p)}^{(m)}+ipa_j\mathbf{A}_{j(l+p)}^{(e)} \right]  \\
\mathbf{A}^\mathrm{l}=\displaystyle\sum_{j=\vert l+p \vert}^{\infty} i^j (2j+1)^{1/2} C_{jlp} \left[ c_j\mathbf{A}_{j(l+p)}^{(m)}+ipd_j\mathbf{A}_{j(l+p)}^{(e)} \right]
\end{array}
\end{equation}
\normalsize
where $\left\lbrace a_j,b_j,c_j,d_j\right\rbrace$ are the Mie coefficients defined in \cite[Chap. 4]{Bohren1983}. These coefficients only depend on the size parameter of the problem ($x=2 \pi R/\lambda$), and both the relative permeability ($\mu_r$) and permittivity ($\epsilon_r$) of the sphere with respect to the surrounding medium. $R$ is the radius of the particle. Note that the multipolar fields in the expansion of $\mathbf{A}^\mathrm{sca}$ and $\mathbf{A}^\mathrm{l}$ contain different radial functions \cite{Bohren1983}. This completes the solution of the problem, as we can now easily compute the value of the fields in all regions. In particular, we can calculate the scattering efficiency, which is a dimensionless measure of the scattered power by the sphere in the far field \cite[Chap. 3]{Bohren1983}. Using the incident beam given by  Eq. (\ref{decomp}), the following expression is obtained for the scattering efficiency:
\begin{equation}
Q^\mathrm{sca}=\sum_{j={ \vert l+p \vert}}^{\infty} \frac{\left( 2j+1 \right)}{x^2} \vert C_{jlp} \vert ^2 \left( \vert a_j \vert ^2 + \vert b_j \vert ^2 \right) 
\label{Csca}
\end{equation}
\begin{figure}[htbp]
\centering\includegraphics[width=12cm]{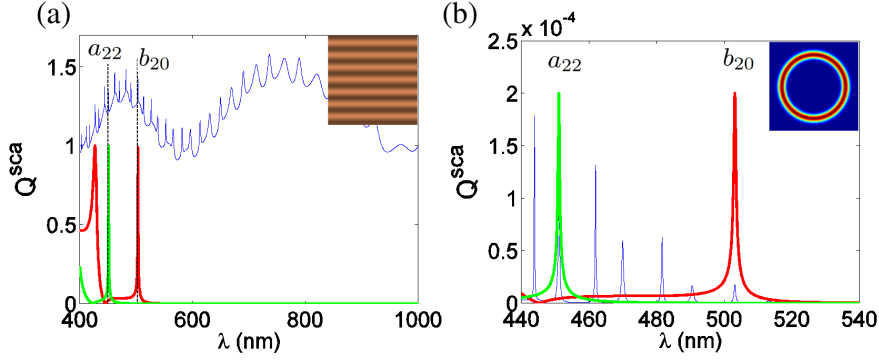}
\caption{Suppression of background in the scattering efficiency ($Q^{sca}$). Input beams are a) plane wave and b) LG$_{0,18}$. In a) no aplanatic lens is used, whereas in b) an aplanatic lens with a NA=0.9 is used. The rest of parameters are kept constant for both plots, \textit{i.e.} $R=1.3 \ \mu$m, $n_r=\sqrt{\epsilon_r \mu_r} = 1.5$, $p=1$. The insets indicate the typical profile of the beam used to excite the sphere. The scattering efficiency is represented with a blue continuous line. The Mie coefficients $b_{20}$ and $a_{22}$ are plotted with a red and a green line respectively. In a) we have indicated with a dashed line the position of two particular resonances for these modes. Note that the ordinate axis in a) and b) are different.} 
\end{figure}
In comparison, the expression of the scattering efficiency for an incident plane wave \cite[Chap. 10]{Jackson1998}, has all $\vert C_{jlp} \vert ^2$ equal to 1 and the sum starts at $j=1$. This means that in conventional Mie Theory, all the Mie coefficients (and consequently all the multipolar modes) have the same relative weight. In our case, the use of an engineered input beam allow us to control the relative weight of the different multipolar modes (given by $a_j$ and $b_j$) with the use of the multipolar content of the incident field (given by $C_{jlp}$). Then, a tightly focused incident beam will allow us to select a few relevant Mie scattering coefficients. On the other hand, the use of beams with a well defined value of $J_z$ allows us to exclude the first $\vert m \vert$ multipolar modes from the scattering cross section.

In order to show these effects, in Fig. 2(a) we plot the conventional Mie scattering efficiency when the incident field is a plane wave, as well as the values of a pair of high order Mie coefficients ($b_{20}$ and $a_{22}$). In this case, the cross section is dominated by the low order multipoles. The higher order resonances appear as little ripples on top of the large oscillation. This is the so-called ripple structure. However, if an engineered cylindrically symmetric beam is used, the lowest $\vert m \vert$ modes can be removed and thus a particular Mie coefficient of interest can be excited. This is shown in Fig. 2(b), where the scattering efficiency associated to a LG$_{0,18}$ beam is depicted, as well as the same two Mie coefficients plotted in Fig. 2(a). 
\begin{figure}[htbp]
\centering\includegraphics[width=12cm]{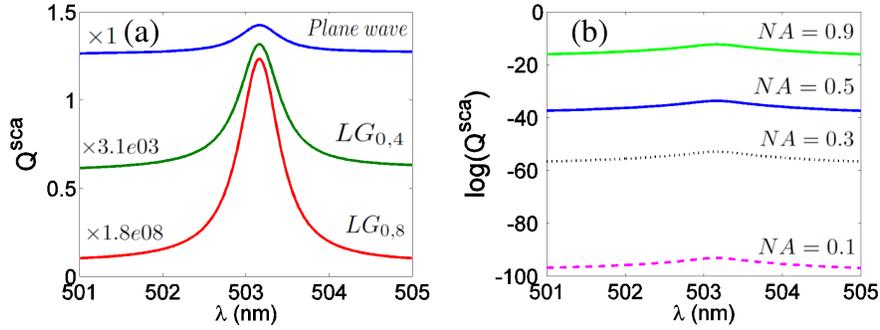}
\caption{Effect of a) the input beam and b) the NA of the lens on the excitation of a single resonance in the scattering efficiency ($Q^{sca}$). In a), three different beams are used: a plane wave and two LG's beams with $l=4,8$ focused with a lens with NA=0.25. In b) we use a LG$_{0,18}$ (same as Fig. 2(b)) but we focus it with different NA=0.1, 0.3, 0.5 and 0.9. The beams always fill the entrance pupil of the lens. Note that in a) the ordinate axis is linear and in b) logarithmic. Also, in a) the efficiency produced by each of the LG beams is multiplied by a different factor so that all the curves are comparable in the same plot. In b) the increase of the floor of the plotted lines is not due to an increase of the background, but rather to an amplification of the tails of the resonances.}
\end{figure}
\begin{figure}[htbp]
\centering\includegraphics[width=12cm]{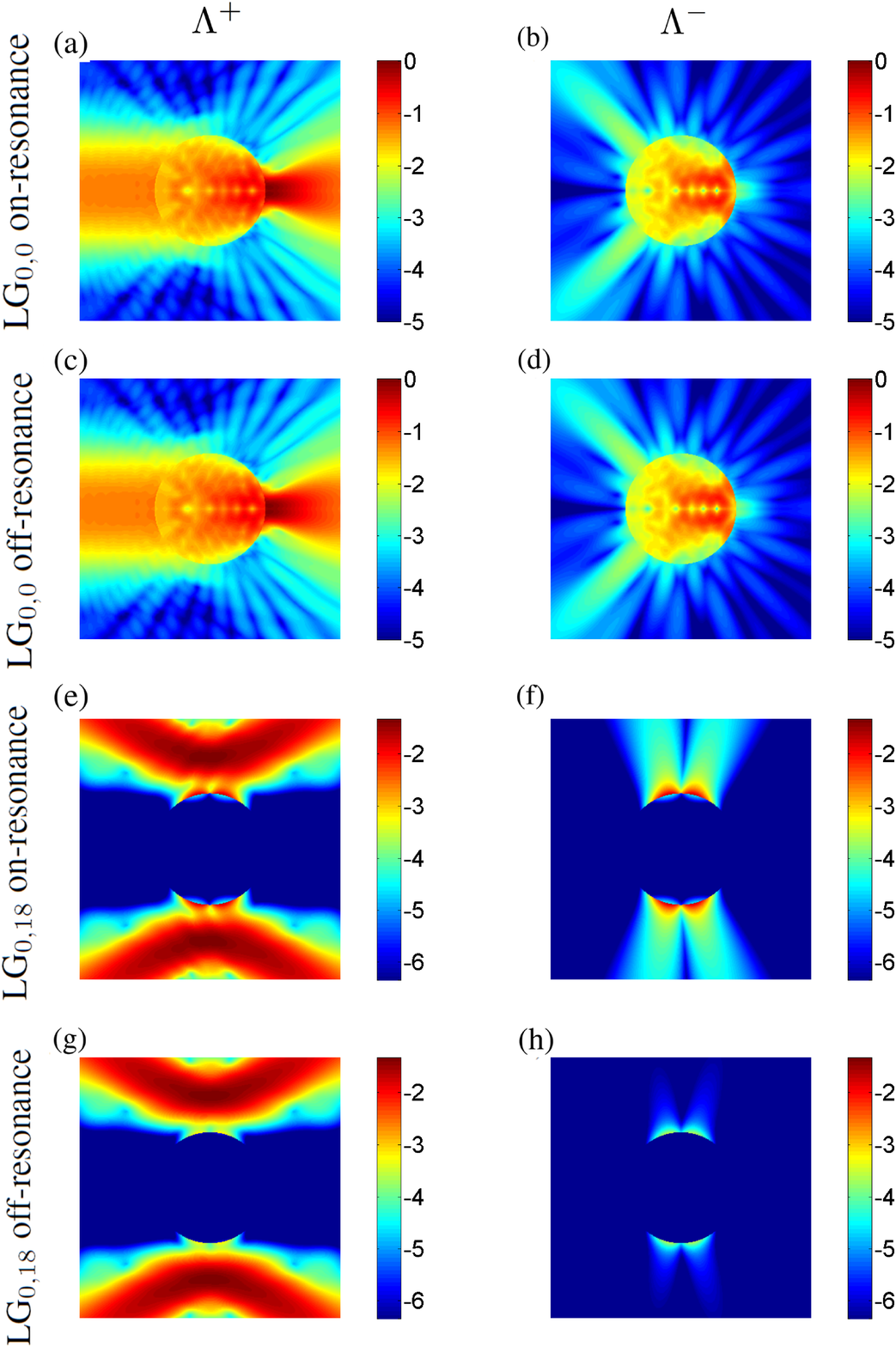}
\caption{Projection into the helicity basis of the intensity of the interior and total field from a sphere with parameters $\left\lbrace R=1.3 \ \mu \mathrm{m},n_r=1.5 \right\rbrace $. The helicity of the incident beam is always $p=1$. The intensity is plotted in logarithmic scale, where 0 corresponds to the maximum of intensity in a).  Images a), b), e) and f) have been simulated for a $\lambda_{\mathrm{resonance}}=503$ nm, whereas a $2$ nm shift has been introduced in the off-resonance wavelength for the others, \textit{i.e.} $\lambda_{\mathrm{off-resonance}} = 505$ nm. The simulations for the LG$_{0,0}$ beam have been executed with a NA=0.25 lens  in order to be able to excite the resonance of order 20. Note that the scattering is dominated by the off-resonance modes. Calculations with higher numerical apertures give similar results. On the contrary, a lens with a NA=0.9 has been used for the excitation with a LG$_{0,18}$. In both cases, the entrance pupil of the lens was filled. Note that, in contrast with b), the mode of order 20 is greatly enhanced when we tune the wavelength to the resonance.}
\end{figure}

One interesting possibility which raises from the ability to freely excite multipolar resonances is that one can address very sharp resonances. The Mie scattering coefficients oscillate from 0 to 1 in a very exotic fashion. This behaviour is given by spherical Bessel functions. Every resonance of a Mie coefficient is given by a zero of a transcendental equation involving Bessel functions of order $j$. These equations have an infinite number of zeros, giving raise to an infinite number of resonances for a given order $j$. In general, it is very difficult to compare the shape of different resonances. Nevertheless, the first peaks of the Mie coefficients do follow a general trend. The higher the order of the Mie coefficient, the sharper the resonance. This implies that the high multipolar resonances of the sphere have smaller linewidths and consequently very large Q factors.  This also means that this sphere will be much more sensitive to the incident radiation. This fact is used in order to excite whispering gallery modes in large spheres, reaching very high Q factors for high values of $j$ \cite{Schiller1991}. In theory, Q could reach values as high as $10^{400}$. Yet, due to radiative losses of the material, the maximum values of Q experimentally achievable are of the order of $Q=10^9$ \cite{Oraevsky2002}.  Although this feature should also be present in the scattering of light from spheres, it is typically hidden due to the background produced by the lower order modes.

As shown in Fig. 3(a), our technique allows to excite a single multipolar resonance with free-space radiation. We fix our attention to the little ripple created by the Mie coefficient $b_{20}$ in the plane wave scattering (also present in Fig. 2(a)). The figure shows how to reduce the background created by the excitation of the lower modes. This has a power drawback though. Increasing the AM of the incident mode reduces the absolute value of the scattering cross section if the NA of the lens is preserved. However, if we increase the NA, the retrieved scattered power increases by many orders of magnitude and the shape of the resonance is maintained. This is shown in Fig. 3(b). Thus, high NA microscope objectives in conjunction with high AM modes can be used to experimentally address single multipolar resonances.

In Fig. 4 we show the interior and the total electric field (scattered plus incident) surrounding the sphere for two different incident beams. Two sets of four plots corresponding to two different incident beams (LG$_{0,0}$ and LG$_{0,18}$) at the same combination of $\left\lbrace R=1.3 \  \mu m,n_r=1.5 \right\rbrace $ are presented. Both incident fields have a well defined helicity $\Lambda^+$ (or $p=1$) in the focus of the lens. Thus, the resonant behaviour of the scattered field will be specially clear when the opposite component of the helicity $\Lambda^-$ is examined, since the contribution due to the incident field is 0. Indeed, if we compare the resonant excitation shown in Fig. 4(f) with the off-resonant behaviour of Fig. 4(h), it can be seen that the intensity differ in two orders of magnitude. Note that as it was shown in Fig. 2, the resonance depicted here is driven by the $b_{20}$ Mie coefficient, hence the single excited multipolar mode is $\mathbf{A}_{20,19}^{(m)}$. Another consequence of the single excitation of a multipolar mode in Fig. 4(e-h) is the fact that in Fig. 4(e-f) the two helicity components of the scattered field have the same power. This is due to the fact that the we are exciting a single magnetic multipole $\mathbf{A}_{20,19}^{(m)}$ and that the multipoles considered in Eq. (\ref{TM-TE}) are a superposition of multipoles with a well defined helicity \cite{preIvan2012PRA}. Finally, it can be observed that unlike Fig. 4(e-h), where the field can be described with a single mode, many multipoles are needed to describe the interferences seen in Fig. 4(a-d). In fact, the pattern-like observed in Fig. 4(a-h) has drawn a lot of attention since 2004, where it was first characterized as photonic nanojet \cite{Chen2004}. Since then, nanojets have been extensively studied and applied in many different fields \cite{Heifetz2009}.

\section{Conclusion}

We have demonstrated how to address multipolar resonances and how to control the scattered field of a spherical object by engineering the incident light. This technique allows to excite single multipolar resonances and also, to choose resonances with very high Q factors even for spheres with sizes of the order of the wavelength of the incident radiation. Besides the fundamental control of material particles that this technique grants, we foresee applications in the fields of cytometry and dark field microscopy, where the control of the AM of the illuminating beams has not been  explored until now. 

\section*{Acknowledgements}
This work was funded by the Australian Research Council Discovery Project DP110103697. G.M.-T. is the recipient of an Australian Research Council Future Fellowship (project number FT110100924). We are thankful to I. Fernandez-Corbaton for some very useful discussions.



\begin{thebibliography}{10}
\newcommand{\enquote}[1]{``#1''}

\bibitem{Gadian1982}
D.~G. Gadian, \emph{Nuclear magnetic resonance and its applications to living
  systems} (Oxford University Press, New York, 1982).

\bibitem{Billah1991}
K.~Y. Billah and R.~H. Scanlan, \enquote{Resonance, tacoma narrows bridge
  failure, and undergraduate physics textbooks,} Am. J. Phys. \textbf{59},
  118--124 (1991).

\bibitem{Garcia-Vidal2010}
F.~J. Garcia-Vidal, L.~Martin-Moreno, T.~W. Ebbesen, and L.~Kuipers,
  \enquote{Light passing through subwavelength apertures,} Rev. Mod. Phys.
  \textbf{82}, 729--787 (2010).

\bibitem{Lopez2002}
R.~Lopez, T.~E. Haynes, L.~A. Boatner, L.~C. Feldman, and J.~R.~F.~Haglund,
  \enquote{Temperature-controlled surface plasmon resonance in vo 2 nanorods,}
  Opt. Lett. \textbf{27}, 1327--1329 (2002).

\bibitem{Mojarad2008}
N.~M. Mojarad, V.~Sandoghdar, and M.~Agio, \enquote{Plasmon spectra of
  nanospheres under a tightly focused beam,} J. Opt. Soc. Am. B \textbf{25},
  651--658 (2008).

\bibitem{preKim2012}
J.~Kim, H.~Son, D.~J. Cho, B.~Geng, W.~Regan, S.~Shi, K.~Kim, A.~Zettl, Y.-R.
  Shen, and G.~Wang, \enquote{Electrical control of plasmon resonance with
  graphene,} arXiv:1206.1124v1 .

\bibitem{Fano1961}
U.~Fano, \enquote{Effects of configuration interaction on intensities and phase
  shifts,} Phys. Rev. \textbf{124}, 1866--1878 (1961).

\bibitem{Fano2010}
B.~Luk'yanchuk, N.~I. Zheludev, S.~A. Maier, N.~J. Halas, P.~Nordlander,
  H.~Giessen, and C.~T. Chong, \enquote{The fano resonance in plasmonic
  nanostructures and metamaterials,} Nat. Mater. \textbf{9}, 707--715
  (2010).

\bibitem{Hergert2012}
W.~Hergert and T.~Wriedt, \emph{The Mie Theory} (Springer, Berlin, 2012).

\bibitem{GLMT_book}
G.~Gouesbet and G.~Gr\'ehan, \emph{Generalized Lorenz-Mie Theories} (Springer,
  Berlin, 2011).

\bibitem{Gouesbet2011}
G.~Gouesbet, J.~Lock, and G.~Gr\'ehan, \enquote{Generalized Lorenz-Mie
  theories and description of electromagnetic arbitrary shaped beams: Localized
  approximations and localized beam models, a review,} J. Quant. Spectrosc. Radiat. Transfer \textbf{112}, 1--27 (2011).

\bibitem{Rose1955}
M.~E. Rose, \emph{Multipole Fields} (Wiley, New York, 1955).

\bibitem{Nora2012}
N.~Tischler, X.~Zambrana-Puyalto, and G.~Molina-Terriza, \enquote{The role of
  angular momentum in the construction of electromagnetic multipolar fields,}
  Eur. J. Phys. \textbf{33}, 1099-1109 (2012).

\bibitem{Allen1992}
L.~Allen, M.~W. Beijersbergen, R.~J.~C. Spreeuw, and J.~P. Woerdman,
  \enquote{Orbital angular momentum of light and the transformation of
  Laguerre-Gaussian laser modes,} Phys. Rev. A \textbf{45}, 8185--8189 (1992).

\bibitem{Sonja2008}
S.~Franke-Arnold, L.~Allen, and M.~Padgett, \enquote{Advances in optical
  angular momentum,} Laser \& Photon. Rev. \textbf{2}, 299--313 (2008).

\bibitem{Lifshitz1982}
V.~B. Berestetskii, L.~P. Pitaevskii, and E.~M. Lifshitz, \emph{{Quantum
  Electrodynamics, Second Edition: Volume 4}} (Butterworth-Heinemann, 1982),
  2nd ed.

\bibitem{preIvan2012PRA}
I.~Fernandez-Corbaton, X.~Zambrana-Puyalto, and G.~Molina-Terriza,
  \enquote{Helicity and angular momentum. a symmetry based framework for the
  study of light-matter interactions.} arXiv:1206.5563v1  (2012).

\bibitem{vandeNes07}
A.~S. van~de Nes and P.~T\"{o}r\"{o}k, \enquote{Rigorous analysis of spheres in
  Gauss-Laguerre beams,} Opt. Express \textbf{15}, 13360--13374 (2007).

\bibitem{Gabi2012OL}
D.~Petrov, N.~Rahuel, G.~Molina-Terriza, and L.~Torner,
  \enquote{Characterization of dielectric spheres by spiral imaging,} Opt.
  Lett. \textbf{37}, 869--871 (2012).

\bibitem{Stilgoe2008}
A.~B. Stilgoe, T.~A. Nieminen, G.~Kn\"{o}ener, N.~R. Heckenberg, and
  H.~Rubinsztein-Dunlop, \enquote{The effect of Mie resonances on trapping in
  optical tweezers,} Opt. Express \textbf{16}, 15039--15051 (2008).

\bibitem{Nieto-Vesperinas2010}
M.~Nieto-Vesperinas, J.~J. S\'{a}enz, R.~G\'{o}mez-Medina, and L.~Chantada,
  \enquote{Optical forces on small magnetodielectric particles,} Opt. Express
  \textbf{18}, 11428--11443 (2010).

\bibitem{Novitsky2011}
A.~Novitsky, C.-W. Qiu, and H.~Wang, \enquote{Single gradientless light beam
  drags particles as tractor beams,} Phys. Rev. Lett. \textbf{107}, 203601
  (2011).

\bibitem{Garcia-Etxarri2011}
A.~Garc\'{i}a-Etxarri, R.~G\'{o}mez-Medina, L.~S. Froufe-P\'{e}rez,
  C.~L\'{o}pez, L.~Chantada, F.~Scheffold, J.~Aizpurua, M.~Nieto-Vesperinas,
  and J.~J. S\'{a}enz, \enquote{Strong magnetic response of submicron silicon
  particles in the infrared,} Opt. Express \textbf{19}, 4815--4826 (2011).

\bibitem{Kuznetsov2012}
A.~I. Kuznetsov, A.~E. Miroshnichenko, Y.~H. Fu, J. Zhang and B.~Luk'yanchuk, \enquote{Magnetic
  light,} Sci. Rep. \textbf{2}, 492 (2012).

\bibitem{Miroshnichenko2012}
A.~E. Miroshnichenko, B.~Luk'yanchuk, S.~A. Maier, and Y.~S. Kivshar,
  \enquote{Optically induced interaction of magnetic moments in hybrid
  metamaterials,} ACS Nano \textbf{6}, 837--842 (2012).

\bibitem{Filonov2012}
D.~S. Filonov, A.~E. Krasnok, A.~P. Slobozhanyuk, P.~V. Kapitanova, E.~A.
  Nenasheva, Y.~S. Kivshar, and P.~A. Belov, \enquote{Experimental verification
  of the concept of all-dielectric nanoantennas,} Appl. Phys. Lett.
  \textbf{100}, 201113 (2012).

\bibitem{Novotny2006}
L.~Novotny and B.~Hecht, \emph{Principles of nano-optics} (Cambridge University
  Press, Cambdrige, MA, 2006).

\bibitem{Gabi2001}
G.~Molina-Terriza, J.~P. Torres, and L.~Torner, \enquote{Management of the
  angular momentum of light: Preparation of photons in multidimensional vector
  states of angular momentum,} Phys. Rev. Lett. \textbf{88}, 013601 (2001).

\bibitem{preIvan2012}
I.~Fernandez-Corbaton, X.~Zambrana-Puyalto, N.~Tischler, A.~Minovich, X.~Vidal,
  M.~L. Juan, and G.~Molina-Terriza, \enquote{Experimental demonstration of
  electromagnetic duality symmetry breaking,} arXiv:1206.0868v1  (2012).

\bibitem{Morse1953}
P.~M. Morse and H.~Feshbach, \emph{Methods of Theoretical Physics} (McGraw-Hill
  and Kogakusha Book Companies, 1953).

\bibitem{Gabi2008}
G.~Molina-Terriza, \enquote{Determination of the total angular momentum of a
  paraxial beam,} Phys. Rev. A \textbf{78}, 053819 (2008).

\bibitem{Rose1957}
M.~E. Rose, \emph{Elementary Theory of Angular Momentum} (Wiley, New York,
  1957).

\bibitem{Tung1985}
W.-K. Tung, \emph{Group Theory in Physics} (World Scientific, Singapore, 1985).

\bibitem{Bohren1983}
C.~F. Bohren and D.~R. Huffman, \emph{Absorption and Scattering of Light by
  Small Particles} (Wiley, New York, 1983).

\bibitem{Jackson1998}
J.~D. Jackson, \emph{Classical Electrodynamics}, vol. 2011 (John Wiley \& Sons,
  New York, 1998).

\bibitem{Schiller1991}
S.~Schiller and R.~L. Byer, \enquote{High-resolution spectroscopy of whispering
  gallery modes in large dielectric spheres,} Opt. Lett. \textbf{16},
  1138--1140 (1991).

\bibitem{Oraevsky2002}
A.~N. Oraevsky, \enquote{Whispering-gallery waves,} Quantum Elec.
  \textbf{32}, 377-400 (2002).

\bibitem{Chen2004}
Z.~Chen, A.~Taflove, and V.~Backman, \enquote{Photonic nanojet enhancement of
  backscattering of light by nanoparticles: a potential novel visible-light
  ultramicroscopy technique,} Opt. Express \textbf{12}, 1214--1220 (2004).

\bibitem{Heifetz2009}
A.~Heifetz, S.~Kong, A.~Sahakian, A.~Taflove, and V.~Backman, \enquote{Photonic
  nanojets,} J. Comput. Theor. Nanosci. \textbf{6}, 1979--1992 (2009).

\end{thebibliography}

\end{document}